\documentclass[prl,showpacs,reprint,longbibliography,superscriptaddress]{revtex4-2}

\usepackage{epsfig,amsmath,amssymb}
\usepackage{xcolor}

\newcommand{\bra}[1]{\langle #1|}
\newcommand{\ket}[1]{|#1\rangle}
\newcommand{\Exp}[1]{\langle#1\rangle}

\begin{document}
	
\title{The Berry phase from the entanglement of future and past light cones: \\
	detecting the timelike Unruh effect}

\author{James Q. Quach} 
\email{quach.james@gmail.com}
\affiliation{Institute for Photonics and Advanced Sensing and School of Physical Sciences, The University of Adelaide, South Australia 5005, Australia}
\affiliation{Commonwealth Scientific and Industrial Research Organisation (CSIRO), Clayton, Victoria 3168, Australia}

\author{Timothy C. Ralph} 
\affiliation{Centre for Quantum Computation and Communication Technology, School of Mathematics and Physics, University of Queensland, Brisbane, Queensland 4072, Australia. }

\author{William J. Munro} 
\affiliation{NTT Basic Research Laboratories \& NTT Research Center for Theoretical Quantum Physics, NTT Corporation, 3-1 Morinosato-Wakamiya, Atsugi-shi, Kanagawa 243-0198, Japan}
\affiliation{National Institute of Informatics, 2-1-2 Hitotsubashi, Chiyoda-ku, Tokyo 101-8430, Japan}


\begin{abstract}
	The Unruh effect can not only arise out of the entanglement between modes of left and right Rindler wedges, but also between modes of future and past light cones. We explore the geometric phase resulting from this timelike entanglement between the future and past, showing that it can be captured in a simple $\Lambda$-system. This provides an alternative paradigm to the Unruh-deWitt detector. The Unruh effect has not been experimentally verified because the accelerations needed to excite a response from Unruh-deWitt detectors are prohibitively large. We demonstrate that a stationary but time-dependent $\Lambda$-system detects the timelike Unruh effect with current technology.
\end{abstract}


\maketitle

The Unruh effect is the intriguing idea that an accelerating observer will view the quantum vacuum as a thermal state~\cite{unruh76}. It arises as a consequence of the theory of relativity applied to quantum mechanics: quantum states are dependent on the spacetime path of the observer. The temperature measured by an accelerating observer however is exceedingly small, requiring a proper acceleration on the order of $10^{20}~\text{ms}^{-2}$ for a temperature of 1 K~\cite{davies75}. Detection of the Unruh effect typically relies on the response of photon detectors. In its most basic form, this is represented by the Unruh-DeWitt detector, which is simply a two-level point monopole~\cite{unruh84}. The Unruh-Dewitt detector has served as the foundational probe detector of relativistic quantum~\cite{davies02,schlicht04,louko06,lin07,brown19,ostapchuk12,salton15,kukita17,martin14,nambu13,steeg09,hu2012} and gravitational fields~\cite{louko08,hodgkinson12,hodgkinson14,ng14,birrell84,langlois06,hodgkinson12}. As such a detector requires energy transfer to excite its response function, it is insensitive to ultraweak fields where excitation events are rare. It is also insensitive to fields in a noisy environments, as its response function is indifferent to noise and signal. This has restricted progress in the detection of the Unruh effect. To make progress in the detection of ultraweak fields or fields in noisy environments, requires a conceptual shift in the fundamental detector paradigm. Here we introduce a probe detector model which does not require energy exchange with the field in which it is measuring. Our detector is a simple $\Lambda$-system with degenerate ground states. The measured field is not probed by an excitation response, but through a geometric or Berry phase response. 

\begin{figure}
	\centering
	\includegraphics[width=.7\columnwidth]{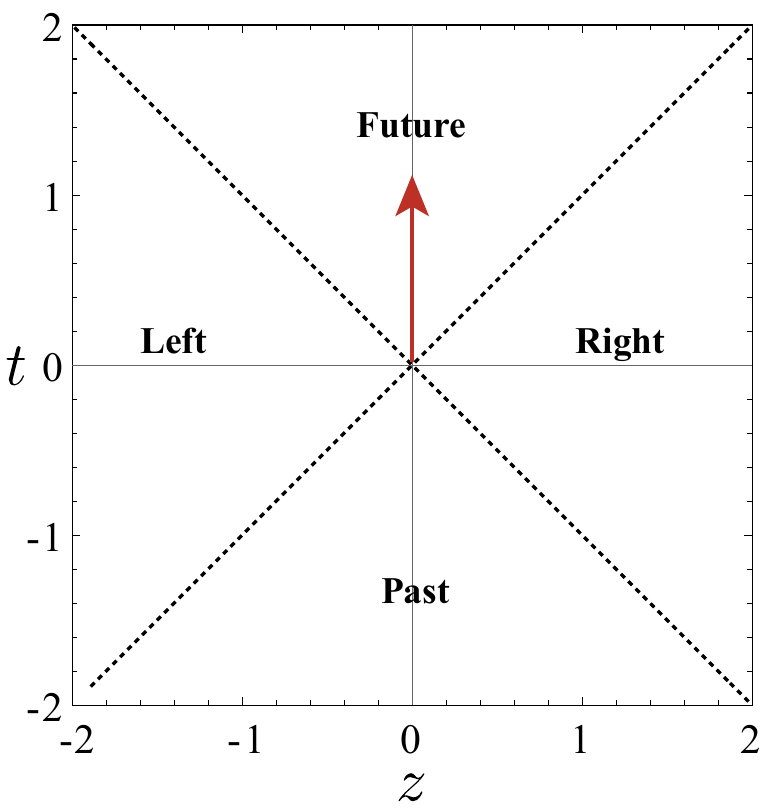}
	\caption{A spacetime diagram separated into four quadrants: the left and right Rindler wedges, and the future and past light cones. The vacuum state can be written as an entangled state between the Rindler wedges, or between the light cones. For an observer in one of these quadrants (e.g. the future), tracing out the unobserved modes (e.g. in the past) leads to the (timelike) Unruh effect. The arrow represents the spacetime trajectory of the detector.} 
	\label{fig:lightcone}
\end{figure}

Previous works have proposed interferometry setups to detect the geometric phase (GP) that result from accelerating atoms~\cite{martin11,hu12}. In these proposals, the atom in one of the arms of the interferometer is accelerated, whilst in the other arm the atom travels inertially. In this setup one must slow down the accelerating atom to precisely match the speed of the inertial atom, in order to eliminate which-path information. This is particularly challenging given that the required acceleration to detect the Unruh effect is on the order of $10^{17}~\text{m/s}^2$. That in itself is problematic, as such large acceleration is likely to change the structure of the atom or ionize it. {We also note a proposal to use muonium atoms in Trojan wave packet states as an Unruh-DeWitt detector~\cite{kalinski05}.  Here we propose an alternative and more practical stationary $\Lambda$-system to measure the Unruh effect by making use of the entanglement between modes in the future and past lightcones.

A uniformly accelerating observer is most conveniently described as a stationary observer in Rindler coordinates~\cite{rindler60}. Here the Unruh effect arises as the result of spacelike entanglement between particles in the left and right Rindler wedges~\cite{unruh76} (Fig.~\ref{fig:lightcone}). Specifically, the vacuum state can be written as an entangled state between two sets of modes spanning the left and right Rindler wedges. As an accelerating observer is confined to just one of these wedges, tracing out the unobserved modes leads to the prediction that such an observer will see a thermalised vacuum. Recently, it has been shown that in theory one could write down the vacuum state similarly as entangled states between modes spanning the future and past light cones~\cite{olson11,higuchi17,ueda21}. If an observer or detector is confined to a spacetime trajectory in one of these cones, tracing out the unobserved modes again will lead to a thermal vacuum states. Here we determine the GP for an observer on one of these trajectories. We demonstrate that the GP can be used to measure the timelike Unruh effect with current technology.

\textit{$\Lambda$-detector.}
The $\Lambda$-detector is a three-level $\Lambda$-system with two ground states as illustrated in Fig.~\ref{fig:lambda}. The ground states ($\ket{g_1}$, $\ket{g_2}$) only couple to the excited state ($\ket{e}$), with transition frequency $\omega/2\pi$. For simplicity we take the ground states to be degenerate. The Hamiltonian of the $\Lambda$-detector interacting with an electromagnetic field in the detector's proper time $\tau$ is $H(\tau)=H_0+H_I(\tau)$, where $H_0=\hbar\omega\ket{e}\bra{e}$ gives the $\lambda$-system's free energy. The interaction Hamiltonian is given by \textcolor{black}{$H_I(\tau)=-q\textbf{r}\cdot\textbf{E}[x(\tau)]\ket{e}(\bra{g_1}+\bra{g_2})+h.c.$}, where $h.c.$ represents the Hermitian conjugate, $qr$ is the electric dipole moment, and \textcolor{black}{$\textbf{E}[x(\tau)]=-\partial \textbf{A}[x(\tau)]/\partial \tau$} is the electric field with \textcolor{black}{\textbf{A} the photon field}. Defining the states $\ket{\pm}\equiv(\ket{g_1}\pm\ket{g_2})/\sqrt{2}$, with corresponding annihilation operators $\sigma_\pm\equiv\ket{e}\bra{\pm}$, the interaction Hamiltonian can be re-written as\textcolor{black}{
\begin{equation}
	H_I(\tau)=-q\textbf{r}\cdot \textbf{E}[x(\tau)](\sigma_++\sigma_+^\dagger)~.
\label{eq:H_I}
\end{equation}}
From this equation, it is clear that $\ket{-}$ is not only an eigenstate, but also a dark state  which does not interact with the electric field. Representing the bright eigenstates subspace as $\ket{p_1}$ and $\ket{p_2}$, we write down the bright component as
\begin{equation}
	\ket{\psi_b(\tau)}=\frac{\sqrt{p_1(\tau)}\ket{p_1(\tau)}+\sqrt{p_2(\tau)}\ket{p_2(\tau)}}{\sqrt{|p_1(\tau)|+|p_2(\tau)|}}~,
\end{equation}
where
\begin{equation}
	\ket{p_i(\tau)}=e^{i\alpha_e(\tau)}\sqrt{p_{i,e}(\tau)}\ket{e}+e^{i\alpha_g(\tau)}\sqrt{p_{i,+}(\tau)}\ket{+}~,
\label{eq:p_i}
\end{equation}
with dynamical phases (DPs) $\alpha(\tau)$. The total system then evolves as 
\begin{equation}
\begin{split}
	\ket{\psi(\tau)}&=e^{i[\beta(\tau)+\alpha_b(\tau)+\phi]}\sqrt{p_1(\tau)+p_2(\tau)}\ket{\psi_b(\tau)}\\
		&\quad+e^{i\alpha_g(\tau)}\sqrt{p_-}\ket{-}~,
\end{split}
\label{eq:psi}
\end{equation}
where $|p_1(\tau)|+|p_2(\tau)|+|p_-|=1$. In Eq.~(\ref{eq:psi}), we make the important observation that only the bright component can pick up a GP $\beta(\tau)$, due to its cyclic interaction with the electric field~\cite{martin11}. Both components also pick-up DPs. We make a further observation that the GP is path-independent in the associated parameter space, and therefore is insensitive to noise \cite{sjoqvist08,berger13}. In contrast, the DP is path-dependent, and sensitive to noise. Phase \textcolor{black}{$\phi=\Delta \mathcal{E}\Delta t$} is an initial phase shift that can be tuned by lifting the energy degeneracy of $\ket{+}$ and $\ket{-}$ (\textcolor{black}{$\Delta \mathcal{E}$}) for a short period of time ($\Delta t$). This can be achieved via a Stark shift of one of the transition frequencies, for example.

\begin{figure}
	\centering
	\includegraphics[width=.9\columnwidth]{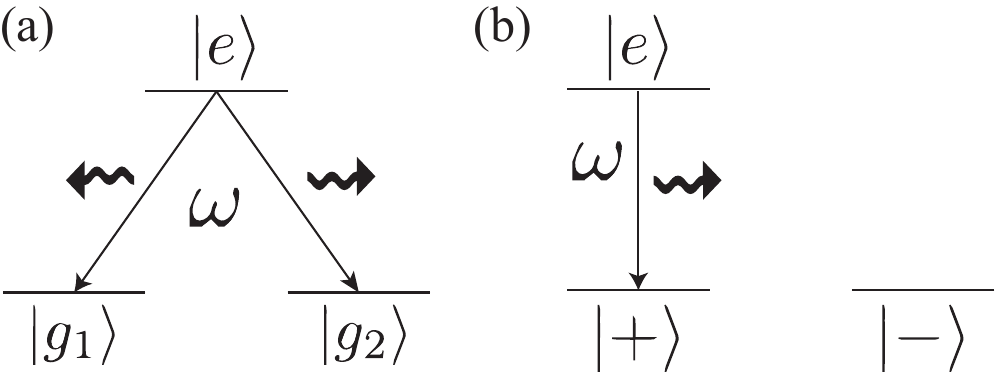}
	\caption{(a) A three-level $\Lambda$-system with two degenerate ground states ($\ket{g_1}$, $\ket{g_2}$) which couple to the excited state ($\ket{e}$), with time-dependent transition \textcolor{black}{angular frequency $\omega(t)$}. (b) The $\Lambda$-configuration can be represented in an L-configuration in the $\ket{\pm}\equiv(\ket{g_1}\pm\ket{g_2})/\sqrt{2}$ basis. Here $\ket{-}$ is a dark state that does not couple to the other states of the system $\{\ket{+},\ket{e}\}$ or the environment. As the $\ket{+}$ does couple to the environment, a relative GP can arise in the system. This GP is used to detect the thermal state of the environment.} 
	\label{fig:lambda}
\end{figure}

The GP $\beta(\tau)$, can be experimentally determined by monitoring the ground state population $\ket{g_1}$ (or $\ket{g_2}$),
\begin{equation}
\begin{split}
	P_1(\tau)&=|\Exp{g_1|\psi(\tau)}|^2\\
		&=\frac{|p_{-}|}{2}+\frac{|p_+(\tau)|}{2}+\sqrt{p_{-}p_+(\tau)}|\cos[\beta(\tau)+\alpha_b(\tau)+\phi]~,
\label{eq:P1}
\end{split}
\end{equation}
where $\sqrt{p_+(\tau)}=\sum^2_{i,j}\sqrt{p_i(\tau)p_{i,+}(\tau)}$. \textcolor{black}{See \textit{Implementation and feasibility} section for further experimental details.}

\textit{Response function in future-past coordinates.}
\label{sec:Olson-Ralph coordinates.}
We have explained how the GP can be experimentally measured in a $\Lambda$-system. Now we explain the theoretical predictions for the GP resulting from the timelike Unruh effect. The future-past (FP) coordinates in the future $(\tau,\zeta)_\text{FP}$ and past light-cones $(\bar{\tau},\bar{\zeta})_\text{FP}$, transform the usual Minkowski coordinate $(t,z)_\text{M}$ as~\cite{olson11}
\begin{align}
	t&=a^{-1}e^{a\tau}\cosh(a\zeta)~, &z&=ca^{-1}e^{a\tau}\sinh(a\zeta)~,\nonumber\\
	t&=-a^{-1}e^{a\bar{\tau}}\cosh(a\bar{\zeta})~, &z&=-ca^{-1}e^{a\bar{\tau}}\sinh(a\bar{\zeta})~,
\label{eq:FP}
\end{align}
where $a$ is a real number in units of s$^{-1}$, while $c$ is the speed of light.

The thermal response of the time-like Unruh effect does not require direct observation of the correlation of the modes of the future and past light cones. As such, we will restrict ourselves to a detector in the future light cone with world line $(\tau,0)_\text{FP}$. For an observer on this world line the Schrodinger equation in FP coordinates is $i\hbar\partial\psi/\partial \tau=H\psi$. In Minkowski coordinates, this corresponds to
\begin{equation}
	i\hbar\frac{\partial\psi}{\partial t}=\frac{H}{at}\psi~.
	\label{eq:schrodinger}
\end{equation} 
The $1/at$ factor is due to the change of variables to Minkowski time. This tells us that a $\Lambda$-detector with energy gap scaled with $1/at$ corresponds to a $\Lambda$-detector on the $(\tau,0)_\text{FP}$ world line~\cite{olson11}.

The two point function of the electric field with respects to Minkowski time [$E(t)=-\partial A(t)/\partial t $] on the detector world line $(\tau,0)_\text{FP}$ is
\begin{equation}
	\begin{split}
		\Exp{E(t[\tau])E(t'[\tau'])}&=-\frac{\hbar}{4\pi^2c^3\epsilon_0}\partial_t\partial_{t'}\frac{1}{(t-t'-i\epsilon)^2}\\
			&=\frac{3\hbar}{32\pi^2c^3\epsilon_0}\frac{a^4e^{-2a(\tau+\tau')}}{\sinh^4[\frac{a}{2}(\tau-\tau'-i\epsilon)]}~.
	\end{split}
\label{eq:EE_t}
\end{equation}
In comparison, the corresponding two point function of the electric field with respects to conformal time  [$E(\tau)=-\partial A(\tau)/\partial \tau $] is
\begin{equation}
	\Exp{E(\tau)E(\tau')}=e^{a(\tau+\tau')}\Exp{E[t(\tau)])E[t'(\tau')])}~.\\
\label{eq:EE_tau}
\end{equation}

\color{black}As dark states do not couple to the environment, only the bright states will contribute to the response function. The bright state $\ket{\psi_b}$ is a two level system in the $\{\ket{e},\ket{+}\}$ basis. Using Eq.~(\ref{eq:EE_t}) and (\ref{eq:EE_tau}) the detector response function is ($\Delta \tau \equiv \tau-\tau'$)
\begin{equation}
\begin{split}
		\mathcal{G}(\omega)&=\frac{q^2}{\hbar^2}|\Exp{e|r|+}|^2\int_{-\infty}^{\infty}d(\Delta\tau)e^{-i\omega\Delta\tau}e^{a(\tau+\tau')}\Exp{E(\tau)E(\tau')}\\
			&=\Gamma(\omega)\Big(1+\frac{a^2}{\omega^2}\Big)\Big(1+\text{coth}\frac{\pi\omega}{a}\Big)
\end{split}
\label{eq:G_w}
\end{equation}
where $\Gamma(\omega)=\frac{\omega^3q^2|\Exp{e|r|+}|^2}{4\pi\epsilon_0\hbar c^3}$ is the spontaneous emission rate. Note the extra exponential factor arises out of a change of variables to conformal time~\cite{olson11}. This response function is similar to the uniformly accelerating ($a'$) case~\cite{hu12}, except $a$ replaces $a'/c$. As such, the thermal response corresponds to temperature $T=\hbar |a|/2\pi k_\text{B}$, where $k_\text{B}$ is the Boltzmann constant. \textcolor{black}{One notes that this Unruh temperature assumes the thermality predicted from theory (as is usual)~\cite{birrell84}, and that a more sophisticated experiment would be needed to confirm this thermality, which is beyond the scope of this paper.}

Interaction with the vacuum leads to a correction in the transition frequency known as a Lamb shift: $\Omega = \omega + \omega_L$. The renormalised correction term is given by~\cite{benatti04}
\begin{equation}
	\omega_L = \frac{i}{2}[\mathcal{K}(-\omega)-\mathcal{K}(\omega)]~,
\end{equation}
with
\begin{equation}
	\mathcal{K}(\lambda)=\text{P}\frac{1}{i\pi}\int^\infty_{-\infty}e^{i\lambda t}\frac{\mathcal{G}(\omega)}{\omega-\lambda}~,
\end{equation}
where P denotes principle value. Using Eq.~(\ref{eq:G_w}) one finds the Lamb shift is of second-order in $\Gamma(\omega)/\omega$. As we work in the $\Gamma(\omega)/\omega<<1$ regime, we can assume the Lamb shift to be negligible.

\textit{Geometric phase.}
\label{sec:Geometric phase.}
Now that we have the response function for a particle on the $(\tau,0)_\text{FP}$ worldline, we can use it to find the eigenstates required to calculate the resulting GP. The evolution of $\rho_b=\ket{\psi_b}\bra{\psi_b}$ is given by the Lindblad master equation
\begin{equation}
	\dot{\rho}_b(\tau)=-i\omega[\sigma_3,\rho_b(\tau)]+\mathcal{G}(\omega)\mathcal{L}[\sigma_+]-\mathcal{G}(-\omega)\mathcal{L}[\sigma_+^\dagger]~,
	\label{eq:lindblad}
\end{equation}
where $\mathcal{L}[O]=O\rho O^\dagger-\frac12\{O^\dagger O,\rho\}$. For initial state $\rho_b(0)=\ket{\psi_b(0)}\bra{\psi_b(0)}$ with $\ket{\psi_b(0)}=\cos\frac{\theta}{2}\ket{e}+\sin\frac{\theta}{2}\ket{+}$, the solution to Eq.~(\ref{eq:lindblad}) is
\begin{equation}
	\rho_b(\tau)=e^{-\delta_+\tau}
	\begin{pmatrix}
		f(\tau)& \frac12e^{-i\omega\tau}\sin\theta\\
		\frac12e^{i\omega\tau}\sin\theta& e^{\delta_+\tau}-f(\tau)~
	\end{pmatrix}~,
	\label{eq:rho}
\end{equation}
where $f(\tau)=e^{-\delta_+\tau}\cos^2\frac{\theta}{2}-(\frac{\delta_-}{\delta_+}-1)\sinh\delta_+\tau$, with $\delta_\pm=\frac12[\mathcal{G}(\omega)\pm \mathcal{G}(-\omega)]$. 

The GP for a mixed state under a non-unitary quasi-cyclic path $\tau_{c}=2\pi/\omega$  is~\cite{tong04}
\begin{equation}
	\beta=\arg\sum_{i=1}^N\sqrt{p_i(0)p_i(\tau_P)}\Exp{p_i(0)|p_i(\tau_P)}e^{-\int_0^{\tau_c}\Exp{p_i(\tau)|\dot{p}_i(\tau)}d\tau}~.
	\label{eq:beta}
\end{equation}
Hence, by diagonalising Eq.~(\ref{eq:rho}) we can get the GP. The eigenvalues of $\rho_b(\tau)$ are
\begin{align}
	p_1(\tau)&=\frac{1}{2}[1+\eta(\tau)]~,\label{eq:eigenvalue1}\\
	p_2(\tau)&=\frac{1}{2}[1-\eta(\tau)]~,\label{eq:eigenvalue2}	
\end{align}
where $\eta(\tau)=\sqrt{\delta^2+e^{-2\delta_+\tau}\sin^2\theta}$ with $\delta=e^{-2\delta_+\tau}\cos\theta+\frac{\delta_-}{\delta_+}(e^{2\delta_+\tau}-1)$. The corresponding eigenstates are
\begin{align}
	\ket{p_1(\tau)}&=\sin\frac{\lambda(\tau)}{2}\ket{e}+e^{i\omega\tau}\cos\frac{\lambda(\tau)}{2}\ket{+}\\
	\ket{p_2(\tau)}&=\cos\frac{\lambda(\tau)}{2}\ket{e}-e^{i\omega\tau}\sin\frac{\lambda(\tau)}{2}\ket{+}
	\label{eq:eigenstate}
\end{align}
where
\begin{equation}
	\tan\frac{\lambda(\tau)}{2}=\sqrt{\frac{\eta(\tau)+\delta(\tau)}{\eta(\tau)-\delta(\tau)}}~.
\end{equation}
As $p_2(0)=0$, only the eigenstate corresponding to $p_1(\tau)$ will contribute to $\beta$: applying Eq.~(\ref{eq:eigenvalue1}) and (\ref{eq:eigenstate}) to (\ref{eq:beta}), the GP is
\begin{equation}
	\beta(a)=\omega\int_0^{\tau_c}\cos^2\frac{\lambda(\tau)}{2}d\tau~.
\label{eq:beta1}
\end{equation}
To first order in $\Gamma(\omega)/\omega$,
\begin{equation}
\begin{split}
	\beta(a)&=\pi(\cos\theta-1)-\frac{\pi^2\Gamma\sin^2\theta}{2\omega^2}\Big(\frac{a^2}{\omega^2}+1\Big)\\
		&\quad\times\Big(\cos\theta+2\coth\frac{\pi\omega}{a}\Big)\text{coth}\frac{\pi\omega}{a}~.
\end{split}
\label{eq:beta2}
\end{equation}

\textcolor{black}{For a stationary detector energy gap, the detector's interactions with the zero point fluctuation of the Minkowski vacuum gives rise to the GP~\cite{chen10}
\begin{equation}
	\beta_0 \equiv  \lim_{a \to 0} \beta(a) = \pi(\cos\theta-1)-\pi^2\frac{\Gamma}{2\omega}(2+\cos\theta)\sin^2\theta~.
\end{equation}}
When $\theta=0$ and $\pi$, which corresponds to $\ket{\psi_b(0)}=\ket{e}$ and $\ket{+}$ respectively, $\Delta\beta(a)\equiv\beta(a)-\beta_0=0$. The GP is maximised near $\theta=\pi/2$. Fig.~(\ref{fig:beta}) plots $\Delta\beta(a)$ for an initial detector state corresponding to $\theta=\pi/2$.

\begin{figure}
	\centering
	\includegraphics[width=.9\columnwidth]{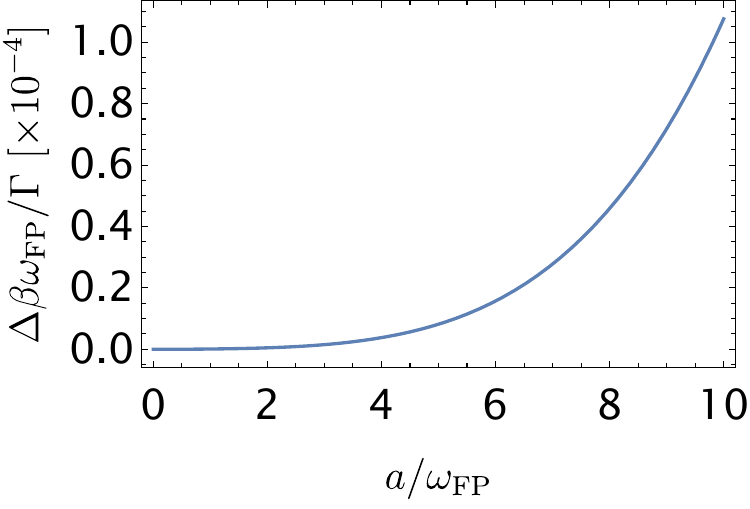}
	\caption{$\Delta\beta$ (normalised to $\Gamma/\omega$) as a function of $a/\omega$, after one quasi-cycle ($\tau_c=2\pi/\omega$).} 
	\label{fig:beta}
\end{figure}

\textit{Sensitivity.} \label{sec:Sensitivity} 
We compare the sensitivity of the $\Lambda$-detector to that of a corresponding Unruh-Dewitt detector. The GP of the $\Lambda$-detector is determined by measuring the groundstate population $P_1$ (or $P_2$).  As we are only interested in the component of the population attributed to the Unruh effect, we define  $\delta P_1(\tau) \equiv P_1(a,\tau)-P_1(0,\tau)$. We choose initial populations to maximise $|\delta P_1(\tau_c)|$: we find that this occurs when the dark state population is $p_-\approx 0.2$. In Fig.~\ref{fig:pop} we plot the shift in groundstate population due to the Unruh effect, $\Delta_{P_1}(\tau)=\delta P_1(\tau)-\delta P_1(0)$.  \textcolor{black}{We see that over ten quasi-cycle $O[\Delta_{P_1}]=0.1$.} In comparison, the sensitivity of the Unruh-Dewitt detector is measured by it's probability of excitation. As the Unruh-Dewitt detector is a two-level system, and noting that the $\Lambda$-system reduces to a two-level system if we do not populate the uncoupled dark state, this probability can be calculated as
\begin{equation}
		P_e(\tau)=|\Exp{e|\psi(\tau)}|^2\\=|p_e(\tau)|~,
\end{equation}
where $\sqrt{p_e(\tau)}=\sum^2_{i,j}\sqrt{p_i(\tau)p_{i,e}(\tau)}$, with initial condition $p_-=0$ with $\theta=\pi$. Under the same conditions as that of the $\Lambda$-detector, $O[P_e]$ never exceeds \textcolor{black}{$10^{-4}$}, meaning that the $\Lambda$-detector is three orders of magnitude more sensitive than the Unruh-Dewitt detector. Moreover, such small signals in the Unruh-DeWitt detector may not be discernable from noise; whereas in the $\Lambda$-detector, the GP is insensitive to noise.

\begin{figure}
	\centering
	\includegraphics[width=.9\columnwidth]{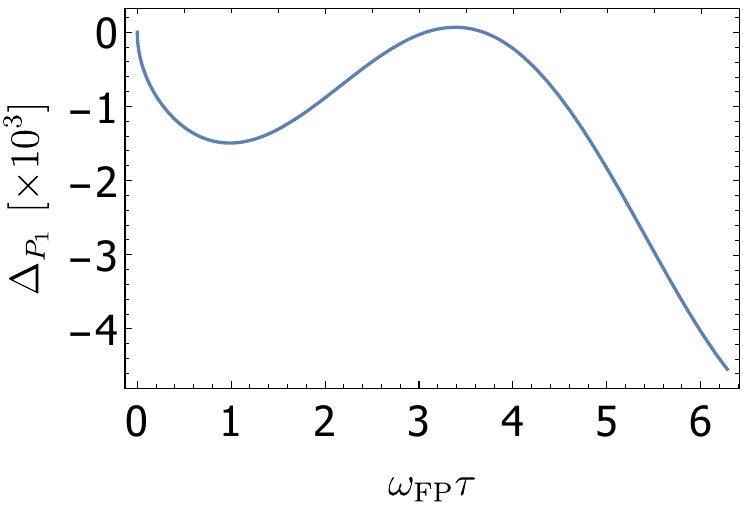}
	\caption{Shift in the groundstate population probability due to the Unruh effect, over \textcolor{black}{ten} quasi-cycles. \textit{Parameters}: $a/\omega=1$, $\Gamma/\omega=10^{-4}$.}
	\label{fig:pop}
\end{figure}

\textit{Implementation and feasibility.} \label{sec:Implementation and feasibility.} 
$\Lambda$-systems are ubiquitous in the selection rules of atoms and molecules, and can be manufactured in artificial-atoms~\cite{buluta011}. Recent work with nitrogen-vacancy (NV) centres has been used to characterise the GP in a $\Lambda$-system~\cite{yale216}. However, the range over which the energy gap can be tuned in atoms may be too restrictive. \textcolor{black}{Instead, we propose a capacitively shunted fluxonium as a possible implementation, which has already been successfully demonstrated as a $\Lambda$-system~\cite{earnest18}. Implementing such a system with gap tunable flux qubits, one may vary the energy gap from  1 to 10 GHz~\cite{zhu10,fedorov10,schwarz13,toida20}. }

\textcolor{black}{As a feasibility case-study we consider a fluxonium implementation with $\omega_M^i/2\pi=5$ GHz. For $n$ quasi-cyclic evolution in FP time, the corresponding Minkowski time is
\begin{equation}
	\ t_\text{c} = \frac{\omega}{a\omega_\text{M}^\text{i}}(e^{na/\omega}-1)~,
	\label{eq:delta_t}
\end{equation}
where $\omega_\text{M}$ is a frequency in Minkowski time $t$. For $a=\omega$, over ten quasi-cycles, $O[t_c]=1~\mu$s. Over this time, the final frequency is $\omega_\text{M}^\text{f}/2\pi=e^{-a/\omega}\omega^\text{i}_\text{M}/2\pi \approx0.2$ MHz. This requires a four-order magnitude change in the energy gap, which is impractical. However, when $\theta=\pi/2$, one observes that $\beta(\tau)$ and $p_+(\tau)$ are independent of the sign of parameter $a$, to first-order in $\Gamma(\omega)/\omega$. This provides a means to conduct the experiment over many cycles without requiring a large change in the energy gap, by alternating between positive and negative $a$. For example, if one switches the sign of $a$ every quasi-cycle, the frequency oscillates between 5.0 GHz and 1.8 GHz.} Under these operating time scales and frequencies, which are well within current technology ranges~\cite{ladd10,raimond01,scully03}, the detector would be detecting an Unruh temperature of 0.6 K. 

\textcolor{black}{Although the GP is independent of the sign of $a$, the DP is not. The DP is given by $\alpha_b(t)=\int_t^{t+\Delta t}\mathcal{H}(t')dt'$, where $\mathcal{H}(t)=H/at$ from Eq.~(\ref{eq:schrodinger}). Therefore, by periodically switching the sign of $a$, the DP can be eliminated whilst the GP accumulates.}

Finally, we note that the large energy gap at small Minkowski time is not physically realisable. In practice, the energy gap may be adiabatically switched on at small time. This would be represented by a switching function, $\chi(\tau)=e^{-\tau^2}$ for example~\cite{olson12}.

\color{black}We summarise the described experimental procedure to determine the GP as follows:
\begin{enumerate}
	\item Adiabatically turn on the energy gap ($\hbar\omega/at$).
	\item Periodically ($\Delta t$) flip the sign of $a$.
	\item Measure the ground population at $P_1(t)$, for some phase shift $\phi$. 	
	\item Initialise the system, but with a different $\phi$.
	\item Repeat steps \textcolor{black}{1 to 4} until maximum $P_1$ is identified. At this point $\beta=\phi$. 
\end{enumerate}
\color{black}

\emph{Conclusion.} \label{Conclusion} 
We have shown how the $\Lambda$-system can be used to detect the timelike Unruh effect that arises out of the entanglement between future and past light cones, with practical operating parameters. This opens the way for experimental verification of the Unruh effect with current technology. More generally, the $\Lambda$-system described here is also a general framework for the detection of ultra-low temperatures. Instead of scaling the energy gap to elicit a thermal response from the vacuum, the GP of a time-independent $\Lambda$-system can detect ambient temperatures, offering a new platform for robust and hypersensitive thermometry.  

\textit{Acknowledgements.} \label{sec:Acknowledgements}
We thank Jay Olson for some useful discussions. JQQ acknowledges financial support for this work from the Ramsay fellowship.

%
%
%

\bibliography{unruh}

\end{document}